\documentclass[12pt]{article}
\begin{document}

\begin{center}
{\large \bf Complexity: An Introduction}
\end{center}

\vspace{0.5in}

\begin{center}
{Rajesh R. Parwani\footnote{Email: parwani@nus.edu.sg}}

\vspace{0.5in}

{University Scholars Programme,\\}
{National University of Singapore,\\}
{Singapore.}

\vspace{0.25in}

{28 January 2002}

%\maketitle
\end{center}

\begin{abstract}
This article summarises a Web-book on ``Complexity" that was developed to introduce undergraduate students to interesting complex systems in the biological, physical and social sciences, and the common tools, principles and concepts used for their study. 

\end{abstract}

\section{Overview}
I use the word {\bf Complexity} to refer to the study of {\bf complex systems}, of which there is no uniformly accepted definition because, well, they are complex. Roughly speaking, one says that a system is complex if it consists of many interacting components (sub-units) and if it exhibits behaviour that is interesting but at the same time not an obvious consequence of the known interaction among the sub-units.

That sounds very vague, especially the use of words like "interesting" and "obvious", but it reflects an evolutionary perspective. For example, a hundred years ago one might have described the study of how a substance changes under heat (phase transitions) as a 
difficult and interesting problem that required one to deal with systems with a large number of interacting components (atoms). However by now very powerful tools, such as thermodynamics and statistical mechancis, have been developed to deal with such {\it equilibrium} systems leading to impressive quantitative agreement between theory and experiment. Though such systems are not commonly referred to as complex, they still provide valuable examples and concepts that have been used in complexity studies. 

Current interest has shifted to {\bf dynamical} systems that are (generally) {\bf out-of-equilibrium} and thus highly {\bf non-linear}. Such sytems actually form the bulk of natural phenomena but for which the theoretical tools are as yet poorly developed. {\bf Some examples of such complex systems or phenomena are: The economy, the stock-market, the weather, ant colonies, earthquakes, traffic jams, living organisms, ecosystems, turbulence, epidemics, the immune system, river networks, land-slides, zebra stripes, sea-shell patterns, and heartbeats.} 
     
There is no single "Theory of Complexity", and it is unlikely that there will ever be one. Rather {\it one hopes that apparently different complex systems can be grouped according to some common features that they have, so that intuition and insight gained in studying one can be transferred to another}. Thus one of the main aims of complexity studies is to develop concepts, principles and tools that allow one to describe features common to varied complex systems. This leads to exciting {\bf interdisciplinary} studies because it turns out that ideas developed to handle complex systems in the physical sciences have relevance also for systems in the biological and social sciences, and {\it vice versa}! 

What are some of the characteristics of complex systems? One often quoted concept is that of {\bf emergence}, which refers to the appearance of laws, patterns or order through the cooperative effects of the sub-units of a complex system. Thus the emergent phenomena or laws are not 
an intrinsic property of the sub-units but rather something that is a property of the system as a whole. Simple examples are those of "temperature" and the "gas laws": At the individual microscopic level, none of those make any sense, but they are features of a large system. More sophisticated examples are of "intelligence" and "conciousness" -- where do they come from ?

Sometimes one sees the phrase "the whole is more than the sum of its parts", as a definition of emergence. This again reflects the non-linearity of the system, whereby the output is not proportional to the input, small changes can give rise to large effects, and the non-obvious results that can be produced in a large system. 

It is important to realise that the universe consists of many hierarchial levels of complexity linked to each other. Each level has  its own emergent patterns and laws: As one goes down from galaxies, solar-systems, planets, ecosystems, organisms, organs, cells, and atoms to quarks, different effective laws emerge. However these laws would not be useful if there was not some degree of {\bf universality}, that is, one hopes that at each level of complexity the same laws apply to varied systems rather than each following its own tune. It is the apparent universality of the laws of physics, for example, that makes the world comprehensible and gives us faith in its ultimate simplicity. For example, at the atomic level weird quantum mechanics rules, but larger systems are well described according to Newtonian laws, while engineers often use empirical rules, and so do the social scientists.

It appears that nature has chosen to be economical (or is that an illusion on our part?), so that the branching of trees or the air-passages in our lungs, the shape of coastlines or clouds, the form of cauliflower or a mountain range, can be described by {\bf fractal} geometry: Such shapes are self-similar over a wide range of scales, thus implying {\bf scale-invariance}, whose hallmark is the appearance of "power-laws". In an  equilibrium system scale-invariance naturally appears at the {\bf critical point} of a {\bf second-order phase transition}, such as that between the liquid and vapour phases of water. However natural systems are out-of-equilibrium and the common appearance of fractals and power-laws in such systems is not as well understaood. {\bf Self-organised criticality} is the idea that many out-of-equilibrium systems naturally organise themselves, without external tuning or prodding, into a state which is at the threshold between complete disorder and complete order: That is, the system arranges itself into a critical state, and so displays scale-invariance and power-laws. 

Living systems are the most complex examples one can think of and it is remarkable how such systems tend in their development towards greater order, organisation and complexity, in contrast to the {\bf arrow of time} dictated by the {\bf Second Law of Thermodynamics}. Of course there is no conflict as the increase in disorder and {\bf entropy} required by the Second Law refers to closed equilibrium systems. Living systems are neither closed nor in equilibrium, but rather use an inflow of energy to drive processes that increase their order (thus decreasing their entropy), and dissipate heat and other waste products that lead to an overall increase in entropy of the universe. One can say that organisms are {\bf dissipative structures}, and have a tendency towards {\bf self-organisation} and {\bf pattern formation}.

Ant-colonies are classic examples of self-organisation. Without a leader (the queen is actually an egg-laying machine) orchestrating everything, and without any of the ants having taken a course in engineering or social science, each ant seems to do its own thing, following a few simple rules that determines its interaction with its environment or its ant-mates. Yet, an incredibly complex and organised society emerges from such an interaction of the many ants. Ant-colonies display remarkable {\bf adaptation} to changing circumstances, using both {\bf feedback} mechanisms and {\bf parallel} analysis of options. In recent years social and computer scientists have taken a keen interest in studying ant colony behaviour in order to help solve problems in their own fields.

Not all systems in nature appear organised or have some pattern to them. Indeed many seem disorderly or ruled by random events. However some of that randomness might only be on the surface. {\bf Chaos} refers to the property of some non-linear dynamical systems whereby they become extremely sensitive to initial conditions and display long-term aperiodic behaviour that seems unpredictable. Though chaotic behaviour might appear essentially random, there is actually hidden order, apparent only in "phase-space" rather in ordinary space. Furthermore, many chaotic systems show universality in their approach to chaos, giving one some predictive power.
Thus discovering that some random-like events are actually chaotic means one has uncovered a simple determinstic basis for the system and so enabled its understandability.  

Often one encounters debates between {\bf reductionism} and {\bf holism}. Reductionists like to get right down to the bottom, meaning they are interested in the basic sub-units that make up the whole and believe that that is where all that is of interest lies, the whole itself being just a complicated and uninteresting consequence of the fundamental laws applied to a large system. In short, knowing the microscopic explains all to the extreme reductionist. Particle physicist are such, and in more recent times some molecular biologists involved with genomics are another example. While it is undoubtedly true that knowledge of the microcomponents of a system and the basic interactions among those is essential for us to progress, it is also a fact that such knowledge by itself is insufficient to predict all the diversity
and novelty that can arise in a large system. (Take for example the task of predicting superconductivity from Schrodinger's equation -- it is a problem that required much effort {\it after} the fact--one knew what to look for. Similarly knowing the whole genome code is not going to predict for us every feature of an organism or a society). 

The problem of precisely deducing the whole (of a large system) from its parts is at least two-fold. Firstly it is a computational problem. Problems with a large number of degrees of freedom are too complicated for exact solutions, and for systems far from equilibrium, as complex systems are, they are also not solvable by the probabilistic averaging methods used for equilibrium systems. In recent years the growth of computer power at low cost has produced the first tool that allows large systems to be simulated or solved numerically. However this brings the second problem: Often one does not have full knowledge of the fundamental dynamics, or the initial conditions, or the problem is still too complicated to be handled directly even by computers.  

Often what is required is some guesswork or intuition to reduce the actual problem to a simpler {\bf model} which can then be tested on a computer. {\bf Computer simulations} of simplified models let one test assumptions quickly, and when the results appear similar to the real world one can take it as {\it plausible} validity of the model and the assumptions. Qualitative similarities of course do not constitute a proof, because other models with different assumptions might give similar results, but at least the insight gained helps one to make further guesses and tests in a particular direction rather than being lost in a mess of detail. In fact one of the most important lessons 
computer simulations have taught us is that a large system with very simple local rules can give rise to collective behaviour
of great complexity and variety, showing on the one hand that complex phenomena need not require complicated rules, but at the same time reminding us how difficult it is (without computers) to deduce the emergent behaviour from the sub-units and their interactions. 

Thus studying the whole is as interesting as studying its parts, as novel structures and emergent laws arise at each level of complexity. The condensed matter physicist studying superconductivity is not going to be replaced by the string theorist, and neither is the ecologist going to be 
become obsolete because of the molecular geneticist. Explaining dynamic patterns, order and emergent laws of a complex system by understanding the organising principles among the sub-units 
is what might be called holism, the counterpoint to reductionism.

\section{Examples}
Let us briefly look at examples to illustrate some of the points above.
These are just apetisers, details are in the main course [\ref{webbook}].

\subsection{Schooling of Fish}
Try out the applet at Ref.[\ref{school}].
Does it not look like a very realistic simulation of fish swimming? The motion of each individual fish is not scripted right from the beginning but rather each individual follows just three simple local rules: cohesion, alignment and separation. Each of the rules is sensible from the biological perspective and so the model is plausible. What is remarkable of course is how the realistic and complicated {\it collective} behaviour emerges from the few simple local rules. There is no leader and none of the individuals has a global plan or perspective, (the motion is not orchestrated from the beginning). 

This example is an example of {\it self-organisation}. Other examples are the herding behaviour of humans, say for example in the stock-market, and the alignment of magnetic spins to form a ferromagnet. Many such examples are studied in [\ref{webbook}].

\subsection{Bacterial Colonies}
Look at the picture of a bacterial colony. It shows a branching structure, which has the property that if one zooms into any region, that part looks similar to the whole. The bacterial colony is an example of a random {\it fractal}. Exact fractals appear the same at different magnification scales while random fractals appear only statistically similar at different magnification scales.

Fractals are ubiquitous in Nature. Another example is the branching network of air-passages in the human lung. The advantages of such a structure are an increase in surface to volume ratio which maximises functional efficiency while minimising material and space costs.

The word "fractal" itself means more than just self-similarity at different scales. It also implies a {\it fractal dimension}.

\subsection{Forest Fires}
Look at the figures which show the number of fires as a function of the area burnt, in different regions of the United States of America and Australia. On a log-log plot one sees that the data is well approximated by straight lines, meaning that the number of fires as a function of area is a power law:

\begin{equation}
N \propto A^{-\alpha}
\end{equation}
with $\alpha \sim 1.3-1.5$. It is important to note that the straight line fits are for a wide range of the data (one can always fit a straight line to a small range), and for different geographical regions. This suggests a {\it universality} in the phenomena that requires a explanation. 

Power laws are observed in many other natural phenomena such as earthquakes and solar flare activity. It has been suggested that these phenomena are examples of {\it self-organised criticality}, that is, the systems are attracted to a state which is between that of total order and  total disorder. The word "critical" is borrowed from well-studied thermal equilibrium systems that undergo second-order phase transitions at critical points and display power laws. However in the case of forest fires (or earthquakes etc.), the systems are far from equilibrium and the power-law behaviour, that is criticality, does not require fine-tuning --- it is self-organised.

Given the complexity fo the actual systems, it is impossible at present to study those systems from first principles. Rather one studies simplified models to check whether power-laws emerge naturally. 

Power laws imply a self-similarity at different scales, so it is natural to suppose that self-organised criticality might be the common dynamical mechanism behind the wide occurence of fractal structures in nature.

Power laws also appear in social contexts, for example the frequency of occurence of words in a literary text, and the magnitude of wars.

\subsection{The Double Pendulum}
Most people are familiar with the simple pendulum: A small heavy object suspended at the end of a thin light string and set into oscillation. For small oscillations (and in the absence of air friction) the motion is that of a "harmonic oscillator": That is, periodic motion with a period proportional to the sqaure-root of the length of the pendulum.

For larger oscillations, the motion of the simple pendulum is still periodic but no longer given by a simple formula. Indeed, for large oscillations the equations governing the motion fo the pendulum are non-linear in contrast to the linear equations for small oscillations. However while the equations are nonlinear, the motion is still regular and predictable.

A double pendulum consists of two simple pendulums in tandem: One attaches a single pendulum to the end of another! The equations of otion are again non-linear for large oscillations but now the motion becomes quite irregular and very sensitive to the initial conditions. This kind of behaviour is the hallmark of {\it chaos}. See the simulation of the double pendulum in the references.

Chaos occurs in many nonlinear systems and it implies that even systems with a few degrees of freedom, and hence naively simple, can show complicated behaviour which is essentially unpredictable on long time scales. However chaos is very different from randomness: The former arises in perfectly deterministic systems while the later is intrinsically nondeterministic, and the distinction between the two at the practical level can be seen by looking at the "phase space" of the system, as we shall see later.

\subsection{The Leopard Spots}
How did the spots on a leopard, the stripes of the zebra or tiger, or the patterns on sea-shells, form? Is there some simpel general framework which can explain these beautiful patterns? Yes, these patterns are called Turing structures, named after the Bristish mathematician who came up with a model to explain such structures. 

Some chemical systems, when maintained far from equilibrium display oscillatory behaviour. When the chemical waves of such a reaction are allowed to diffuse through a medium at different rates andthe resulting pattern stabilised, one obtains Turing structures.

Many other interesting far from equilibrium systems show cyclical behaviour.

\section{Summary}

The aim of the {\bf Complexity} course in Ref.[\ref{webbook}] is two fold:\\

(1) To provide the student with a relatively gentle introduction to the concepts mentioned above so that they can continue on their own in greater depth.

(2) To broaden the students  horizons by introducing  to them the  interdisciplinary nature of complex systems studies, which exemplifies in a concrete sense the often quoted "unity of knowledge".\\

As in all popular fields, one often finds in descriptions of "Complexity" misleading or hyped statements, and metaphorical deviations. So some caution and cynicism is required in filtering the raw data from a search, especially over the Internet. 

In conclusion, explore the Web-book [\ref{webbook}], the links and exercises therein,  and hope for some enlightenment.

\section{References} 

\begin{enumerate}

\item Complexity: A Web-book at http://staff.science.nus.edu.sg/$\sim$parwani/complexity.html \label{webbook}

\item The dreams of reason : the computer and the rise of the sciences of complexity, by Heinz R. Pagels. 

\item The Quark and The Jaguar, by M. Gell-Mann.

\item Web of Life, by F. Capra

\item How Nature Works, by P. Bak

\item The Self-Made Tapestry, by P. Ball

\item The Computational Beauty of Nature, by G. Flake 

\item Butterfly Economics, by P. Omerod 

\item A Fish schooling applet at \\
http://www.codepuppies.com/$\sim$steve/aqua.html \label{school}

\item The double pendulum applet at \\
http://www.maths.tcd.ie/$\sim$plynch/SwingingSpring/doublependulum.html
\end{enumerate}

\end{document}